\documentclass[letterpaper]{JHEP3}
\usepackage{amsmath}
\usepackage{dsfont}
\usepackage{bm}

\title{ Thermodynamics of Lovelock black holes with a nonminimal scalar field}

\author{Francisco Correa\\
Centro de Estudios Cient\'{\i}ficos (CECs), Valdivia, Chile.\\
E-mail: \email{correa-at-cecs.cl}}

\author{Mokhtar Hassaine\\
Instituto de Matem\'atica y F\'{\i}sica, Universidad de Talca,\\
Casilla 747, Talca, Chile.\\
E-mail: \email{hassaine-at-inst-mat.utalca.cl}}

\abstract{We source the Lovelock gravity theories indexed by an integer $k$ and fixed by requiring a unique anti-de Sitter vacuum with a self-interacting nonminimal
scalar field in arbitrary dimension $d$. For
each inequivalent Lovelock gravity theory indexed by the integer $k$, we establish the existence of a two-parametric self-interacting potential that permits to
derive a class of black hole solutions with planar horizon for any arbitrary value of the nonminimal coupling parameter.   In the thermodynamical analysis of the solution, we show that, once regularized the Euclidean action, the mass contribution coming form the
gravity side exactly cancels, order by order, the one arising from the matter part yielding to a vanishing mass. This result is in accordance with the fact that the entropy of the solution, being proportional to the lapse function evaluated at the horizon, also vanishes. Consequently, the integration constant appearing in the solution is interpreted as a sort of hair which turns out to  vanish at high temperature. }

\begin{document}

\section{Introduction}
During the last decades, the interests on higher-dimensional physics have grown up particulary concerning
the higher-dimensional General Relativity. In this context, apart from the standard Einstein-Hilbert action, there
also exists an interesting gravity theory in dimension $d\geq 5$ involving higher  powers of
the curvatures such that the field equations for the metric are at most of second-order. This
theory known as the Lovelock gravity has been first implemented in five dimensions by
Lanczos \cite{LAN} and then
generalized in higher dimension $d\geq 5$ by Lovelock \cite{LOV}. The
resulting action is a
$d-$form constructed out of the vielbein, the spin connection and
their exterior derivative without using the Hodge dual. The invariance under local Lorentz transformations of the
Lovelock Lagrangian can be extended to a local anti-de Sitter (AdS) or
Poincar\'e symmetry through a particular choice of the coefficients
appearing in the Lovelock expansion. In both cases, the resulting
Lagrangian is a Chern-Simons form whose supersymmetric extensions are also known; see
\cite{Zanelli:2005sa} for a good review on Chern-Simons
(super)gravity. The Lovelock gravity or its Chern-Simons particular
case have been shown to possess (topological) AdS black hole
solutions with interesting thermodynamical properties
\cite{Banados:1993ur,Cai:1998vy,Crisostomo:2000bb,Aros:2000ij}
generalizing those obtained in the Einstein-Gauss-Bonnet case
\cite{Boulware:1985wk,Cai:2001dz}.

In order to source the Lovelock gravity with the purpose of obtaining black hole
solutions, nonminimally
coupled scalar field can be an excellent candidate. Indeed, the nonminimal coupling may be useful
to escape the
standard no-hair theorems, and black hole solutions are known in this case in standard Einstein gravity
(with and without cosmological constant) in four dimensions with a conformal coupling
\cite{Bekenstein:1974sf,Bocharova:1970,Martinez:2002ru,Martinez:2005di}. Recently, there has been a renewal interest
concerning this kind of source particularly in order to gain a better
understanding of some unconventional superconductors
\cite{Horowitz:2010gk,Hartnoll:2008kx,Herzog:2009xv}. Indeed, it is believed that black
hole solutions  with scalar hair at low temperature that
disappears at high temperature will be of particular importance for the unconventional superconductors since they will reproduce
the correct behavior of the superconductor phase diagram.

In the present work, we consider a series of particular Lovelock gravity actions indexed by an integer $k$
in dimension $d$, and sourced by a self-interacting nonminimal
scalar field. To be more precise, the arbitrary coefficients appearing in the Lovelock series
are fixed by requiring that the resulting theory has a unique anti-de Sitter vacuum with a fixed
cosmological constant  This yields to $k=1,2\cdots, [\frac{d-1}{2}]$ inequivalent gravity theories where $k=1$ corresponds
to the standard Einstein-Hilbert action. Note that such model with a mass term potential has been considered in \cite{Gaete:2013ixa,Gaete:2013oda}, and black hole
solutions with planar horizon have been obtained for two particular values of the nonminimal
coupling parameter. Here, we generalize these solutions and obtain black hole solutions for any
value of the nonminimal coupling parameter. This
task is achieved thanks to the introduction of a much more general self-interacting term that depends explicitly on the integer index $k$. More specifically, we show that for
each inequivalent Lovelock gravity theory indexed by $k$, there exists a judicious choice for the self-interacting potential that permits to
derive black hole solutions for arbitrary value of the nonminimal coupling parameter. The solutions are shown to be uniparametric and reduce to those
derived in  \cite{Gaete:2013ixa,Gaete:2013oda} for the special values of the nonminimal coupling parameter. The thermodynamics study of the
solutions realized  through the Hamiltonian analysis reveals that these configurations have a zero mass as well as a vanishing entropy. More precisely,
we show that the mass contribution  coming from the gravity side exactly cancels the mass contribution inherent to the matter source.  The vanishing of the entropy can be also corroborated by the fact that the entropy of the solutions is proportional to the lapse function evaluated at the horizon. Since the constant appearing in the black hole solution does  not contribute to any conserved charge, we interpret it as a sort of hair which turns out to be inversely proportional to the temperature. Hence, at high temperature, this hair will disappear and this kind of behavior is excepted in the unconventional superconductors in order to correctly reproduce the phase diagram.

The plan of the paper is organized as follows. In the next section, we present the model, the field equations and the solutions. Sec. III is devoted to the thermodynamical analysis while our conclusions and comments are given in the last section.

\section{Lovelock black hole solution with arbitrary nonminimal coupling parameter}
We  consider a generalization of the Einstein-Hilbert  gravity
action in arbitrary dimension $d$ yielding at most to second-order
field equations for the metric and known as the Lovelock Lagrangian.
This latter is a $d-$form constructed with the vielbein $e^a$, the
spin connection $\omega^{ab}$, and their exterior derivatives
without using the Hodge dual. The Lovelock action is a polynomial of
degree $[d/2]$ (where $[x]$ denotes the integer part of $x$) in the
curvature two-form, $R^{ab} = d\,\omega^{ab} + \omega^{a}_{\;c}
\wedge \omega^{cb}$ as
\begin{subequations}
\begin{eqnarray}
&&\int \sum_{p=0}^{[d/2]}\alpha_p~ L^{(p)},\\
&& L^{(p)}=\epsilon_{a_1\cdots a_d} R^{a_1a_2}\cdots
R^{a_{2p-1}a_{2p}}e^{a_{2p+1}}\cdots  e^{a_d},
\end{eqnarray}
\end{subequations}
where the $\alpha_p$ are arbitrary dimensionful coupling constants
and where wedge products between forms are understood. Here
$L^{(0)}$ and $L^{(1)}$ are proportional respectively to the
cosmological term and the Einstein-Hilbert Lagrangian. As shown
in Ref. \cite{Crisostomo:2000bb}, requiring the Lovelock  action to
have a unique AdS vacuum with a unique cosmological constant, fixes
the $\alpha_p$  yielding to a series of actions indexed by an
integer $k$, and given by
\begin{eqnarray}
I_k=-\frac{1}{2k(d-3)!}\int \sum_{p=0}^{k}\frac{C^k_p}{(d-2p)}~
L^{(p)},\qquad\qquad 1\leq k\leq \Big[\frac{d-1}{2}\Big], \label{Ik}
\end{eqnarray}
where $C^k_p$ corresponds to the combinatorial factor. The global
factor in front of the integral is chosen such that the gravity
action (\ref{Ik}) can be re-written in the standard fashion as
\begin{eqnarray}
I_k=\frac{1}2\int d^{d}x\,\sqrt{-g} \Big[&&R+\frac{(d-1)(d-2)}{k}+\frac{(k-1)}{2(d-3)(d-4)}{\cal L}_{GB}\nonumber\\
&&+
\frac{(k-1)(k-2)}{3!(d-3)(d-4)(d-5)(d-6)}{\cal L}_{(3)}+\cdots
\Big], \label{Ik2}
\end{eqnarray}
where ${\cal L}_{GB}=R^{2}-4\,R_{\mu \nu}R^{\mu
\nu}+R_{\alpha\beta\mu\nu}R^{\alpha\beta\mu\nu}$ stands for the
Gauss-Bonnet Lagrangian, and  ${\cal L}_{(3)}$ is given by
\begin{eqnarray*}
{\cal L}_{(3)}&=&R^3   -12RR_{\mu \nu } R^{\mu \nu } + 16\,R_{\mu
\nu }R^{\mu }_{\phantom{\mu} \rho }R^{\nu \rho } + 24 R_{\mu \nu
}R_{\rho \sigma }R^{\mu \rho \nu \sigma }+ 3RR_{\mu \nu \rho \sigma
} R^{\mu \nu \rho \sigma }
\nonumber \\
&&-24R_{\mu \nu }R^{\mu} _{\phantom{\mu} \rho \sigma \kappa } R^{\nu
\rho \sigma \kappa  }+ 4 R_{\mu \nu \rho \sigma }R^{\mu \nu \eta
\zeta } R^{\rho \sigma }_{\phantom{\rho \sigma} \eta \zeta }-8R_{\mu
\rho \nu \sigma } R^{\mu \phantom{\eta} \nu \phantom{\zeta}
}_{\phantom{\mu} \eta \phantom{\nu} \zeta } R^{\rho \eta \sigma
\zeta }.
\end{eqnarray*}
Note
that in odd dimension $d=2n-1$ and for $k=n-1$, the corresponding
action $I_{n-1}$ is a Chern-Simons action, that is a $(2n-1)-$form
whose exterior derivative can be written as the contraction of an
invariant tensor with the wedge product of $n$ curvatures two-forms.
In even dimension $d=2n$, the maximal value of $k$ is $n-1$, and in
this case the resulting gravity action has a Born-Infeld like
structure since it can be written as the Pfaffian of the $2-$form
$\bar{R}^{ab}=R^{ab}+e^ae^b$.  The gravity theories $I_k$ have been
shown to possess black hole solutions with interesting features, in
particular concerning their thermodynamics properties, see
\cite{Crisostomo:2000bb} and \cite{Aros:2000ij}.

For each $k\geq 2$, we source the gravity actions $I_k$  with a self-interacting and nonminimally coupled scalar field, that is
\begin{align}
S_k=&I_k-\int d^{d}x\,\sqrt{-g} \Big[
\frac{1}{2}\partial_{\mu}\Phi\partial^{\mu}\Phi+\frac{\xi}2 R\Phi^2
+U_k(\Phi)\Big],\label{Sk}
\end{align}
where $\xi$ denotes the nonminimal coupling parameter and $U_k$ is a potential term which depends explicitly on the index $k$, and whose
form will be given below. The field equations read
\begin{subequations}
\label{eqsmotionk}
\begin{eqnarray}
&&{\cal G}^{(k)}_{\mu\nu}=T_{\mu\nu},\\
&&\Box \Phi = \xi R \Phi+\frac{d U_k}{d\Phi},
\end{eqnarray}
\end{subequations}
where ${\cal G}^{(k)}_{\mu\nu}$ is the gravity tensor associated to
the variation of the action $I_k$ (\ref{Ik}),
\begin{eqnarray*}
{\cal G}^{(k)}_{\mu\nu}=&&G_{\mu\nu}-\frac{(d-1)(d-2)}{2k}g_{\mu\nu}+\frac{(k-1)}{2(d-3)(d-4)}K_{\mu\nu}\\
+&&\frac{(k-1)(k-2)}{3!(d-3)(d-4)(d-5)(d-6)}S_{\mu\nu}+\cdots
\end{eqnarray*}
where $K_{\mu\nu}$ is the Gauss-Bonnet tensor
$$
K_{\mu\nu}=
2\big(RR_{\mu\nu}-2R_{\mu\rho}R^{\rho}_{\phantom{\rho}\nu}-2R^{\rho\sigma}R_{\mu\rho\nu\sigma}
+R_{\mu}^{\phantom{\mu}\rho\sigma\gamma}R_{\nu\rho\sigma\gamma}\big)
-\frac{1}{2}\, g_{\mu\nu}\mathcal{L}_{GB}
$$
and $S_{\mu\nu}$ arises from the variation of ${\cal L}_{(3)}$,
\begin{eqnarray*}
S_{\mu\nu}&=&\,3\Big(R^2 R_{\mu \nu } - 4 R R_{\rho  \mu } R^{\rho}
_{\phantom{\rho} \nu } - 4R^{\rho \sigma }R_{\rho \sigma }R_{\mu \nu
} +8 R^{\rho \sigma }R_{\rho  \mu }R_{\sigma  \nu } - 4 R R^{\rho
\sigma }R_{\rho \mu  \sigma  \nu }
\nonumber \\
&&  +8 R^{\rho \kappa }R^\sigma _{\phantom{\sigma }\kappa }R_{\rho
\mu  \sigma  \nu } -16 R^{\rho \sigma }R^\kappa _{\phantom{\kappa }
(\mu }R_{ |\kappa \sigma \rho |  \nu ) } + 2 R R^{\rho \sigma \kappa
}_{\phantom{\rho \sigma \kappa }\mu  }R_{\rho \sigma \kappa  \nu }
+R_{\mu \nu }R^{\rho \sigma \kappa \eta } R_{\rho \sigma \kappa \eta
}
\nonumber \\
&&- 8 R^\rho _{\phantom{\rho }(\mu }R^{\sigma \kappa \eta
}_{\phantom{\sigma \kappa \eta } |\rho | }R_{|\sigma \kappa \eta |
\nu ) } - 4 R^{\rho \sigma }R^{\kappa \eta }_{\phantom{\kappa \eta }
\rho \mu }R_{\kappa \eta  \sigma  \nu }+8R_{\rho \sigma }R^{\rho
\kappa  \sigma  \eta }R_{\kappa  \mu  \eta  \nu } - 8 R_{\rho \sigma
}R^{\rho  \kappa \eta }_{\phantom{\rho  \kappa \eta }\mu }R^\sigma
_{\phantom{\sigma } \kappa \eta  \nu }
\nonumber \\
&&+4 R^{\rho \sigma  \kappa \eta }R_{\rho \sigma  \zeta  \mu
}R_{\kappa \eta  \phantom{\zeta } \nu }^{\phantom{\kappa \eta }\zeta
} - 8 R^{\rho  \kappa  \sigma  \eta }R^\zeta _{\phantom{\zeta }\rho
\sigma  \mu }R_{\zeta  \kappa \eta  \nu }- 4R^{\rho \sigma \kappa
}_{\phantom{\rho \sigma \kappa } \eta  } R_{\rho \sigma \kappa \zeta
}R^{\eta  \phantom{\mu }\zeta }_{\phantom{\eta } \mu \phantom{\zeta
} \nu }\Big)-  \frac12 g_{\mu \nu }{\cal L}_{(3)}.
\end{eqnarray*}
In the matter part of the equations, $T_{\mu\nu}$
stands for the energy-momentum tensor of the scalar field whose expression is given by
\begin{eqnarray}
\label{tmunusf} T_{\mu \nu}=&&\partial_{\mu}\Phi\partial_{\nu}
\Phi-g_{\mu
\nu}\Big(\frac{1}{2}\,\partial_{\sigma}\Phi\partial^{\sigma}\Phi+U_k(\Phi)\Big)+\xi\left(g_{\mu
\nu}\Box-\nabla_{\mu}\nabla_{\nu}+G_{\mu\nu}\right)\Phi^{2}.
\end{eqnarray}
For each inequivalent theory $k\geq 2$ and for $\xi\not=\frac{1}{4}$, the potential is given by the sum of the six following terms
\begin{eqnarray}\label{potLov}
&&U_k(\Phi)=\frac{1}{(4\xi-1)^2}\left[\alpha_1\Phi^2+\alpha_2\,b\Phi^{\frac{1}{2\xi}}+\alpha_3\,b^2\Phi^{\frac{1-2\xi}{\xi}}+\alpha_4\Phi^{\frac{2k}{k-1}}
+\alpha_5\,b\Phi^{\frac{4\xi+k-1}{2\xi(k-1)}}+\alpha_6\,b^2\Phi^{\frac{2\xi(2-k)+k-1}{\xi(k-1)}}\right]\nonumber\\
\end{eqnarray}
and depends of a parameter $b$ and the constants $\alpha_i$ which read
\begin{eqnarray*}
&&\alpha_1=8\xi d(d-1)(\xi-\xi_d)(\xi-\xi_{d+1}),\qquad \alpha_2=-16\xi^2 (d-1)(\xi-\xi_d),\qquad \alpha_3=2\xi^2\\
&&\alpha_4=-\frac{8\xi^{\frac{k}{k-1}}(1+(k-1)d)((k-1)d+2-k)}{(k-1)k}\left(\xi-\hat{\xi}_{k,d}\right)
\left(\xi-\hat{\xi}_{k,d+1}\right)\\
&&\alpha_5=\frac{16\xi\left(k(d-1)-(d-2)\right)-4d(k-1)+8(k-1)}{(k-1)}\xi^{\frac{2k-1}{k-1}},\qquad\alpha_6=-\frac{8\xi+2(k-1)}{k-1}\xi^{\frac{2k-1}{k-1}}
\end{eqnarray*}
Here $\xi_{d}$ denotes the conformal coupling in $d$ dimensions and we have defined $\hat{\xi}_{k,d}$  by
\begin{eqnarray}
\label{xi}
\xi_d=\frac{d-2}{4(d-1)},\qquad \hat{\xi}_{k,d}=\frac{(d-2)(k-1)}{4\left[(d-1)k-(d-2)\right]}.
\end{eqnarray}
Various comments can be made concerning the particular form of the potential (\ref{potLov}). We anticipate that
this potential naturally emerges looking for solutions of the field equations with an Ansatz of the form
\begin{eqnarray}
ds^2=-F(r)dt^2+\frac{dr^2}{F(r)}+r^2d\vec{x}_{d-2}^2,\qquad \Phi=\Phi(r).
\label{ansatz}
\end{eqnarray}
We also note that the self-interacting term
depends explicitly on a coupling constant $b$ as well as on the index parameter of the Lovelock theories $k$, and this dependence on $k$
only concerns the last three terms of the potential. The first three terms of $U_k$  exactly reproduce the potential that appears in many different contexts involving nonminimally coupled scalar field as for examples for the stealth configuration \cite{AyonBeato:2004ig} on the BTZ black hole \cite{Banados:1992wn}, or when looking for AdS wave backgrounds \cite{AyonBeato:2005qq}, or for black hole solutions when adding axionic fields coupled with the scalar field \cite{Caldarelli:2013gqa}. For $b=0$ and $\xi=\hat{\xi}_{k,d}$ or $\xi=\hat{\xi}_{k,d+1}$,
the potential reduces to the mass term considered in \cite{Gaete:2013ixa,Gaete:2013oda}  for which black hole solutions have been found. It is also clear from its expression that $U_k$ is not well defined for $\xi=\frac{1}{4}$, and its derivation for this particular coupling involves
logarithmic pieces.

Let us now present some solutions of the field equations (\ref{eqsmotionk}). For $k\geq 2$ and $\xi\not=\frac1{4}$, an uniparametric  black hole solution
with planar horizon is given by
\begin{eqnarray}
\label{solk}
&&ds^{2}=- r^2\left(1-\left[\xi(ar+b)^{\frac{4\xi}{4\xi-1}}\right]^{\frac{1}{k-1}}\right)dt^{2}+\frac{dr^2}{ r^2\left(1-\left[\xi(ar+b)^{\frac{4\xi}{4\xi-1}}\right]^{\frac{1}{k-1}}\right)}+{r^{2}}d\vec{x}_{d-2}^2,\nonumber\\
&&\Phi(r)=(ar+b)^{\frac{2\xi}{4\xi-1}},
\end{eqnarray}
where $a$ is an integration constant. We first stress that this solution is valid for any arbitrary value of the nonminimal
coupling parameter $\xi\not=0$, and for $\xi\in ]0,\frac{1}{4}[$, the solution is asymptotically AdS
while for $\xi>\frac{1}{4}$ the asymptotic behavior of the metric is faster that the usual AdS one. It is interesting to note that
the singularity $r_s$ is localized at $r_s=-\frac{b}{a}$ but it can always be hidden by the horizon $r_h=-\frac{b}{a}+\frac{\xi^{\frac{1-4\xi}{4\xi}}}{a}$ provided that the constant $a>0$. For $b=0$, and for $\xi=\hat{\xi}_{k,d}$ or for $\xi=\hat{\xi}_{k,d+1}$, the solutions (\ref{solk})
reduce to those found in the case of a mass term potential \cite{Gaete:2013ixa,Gaete:2013oda}. We would like also to point out  an important remark concerning the lapse function appearing in the metric solution
that will be of importance in the thermodynamics study of these solutions. The metric function solution
can be expressed in term of the scalar field as
$$
r^2\left(1-\left[\xi(ar+b)^{\frac{4\xi}{4\xi-1}}\right]^{\frac{1}{k-1}}\right)=r^2\left(1-(\xi\Phi^2)^{\frac{1}{k-1}}\right)
$$
and, hence the localization of the horizon $r_h$ is such that
\begin{eqnarray}
\left(1-\xi\Phi^2\right)|_{r_h}=0.
\label{truc}
\end{eqnarray}
However, as we will see below, the entropy of the solutions is always proportional to this quantity (\ref{truc}), and consequently the solutions (\ref{solk})
will have a zero entropy. We also remark that the existence of these solutions is strongly inherent to the presence of the higher curvature terms present in the
Lovelock Lagrangian. Indeed, it is clear from the different expressions obtained here that the standard GR limit $k=1$ is singular, and hence these solutions
are only effective for the higher order Lovelock terms $k\geq 2$.

To conclude this section, we present the solution for the coupling $\xi=\frac{1}{4}$. The self-interacting potential is given by
\begin{eqnarray}
&&U_{[k,\xi=\frac{1}{4}]}(\Phi)=\frac{1}{8k(k-1)}\Big\{4\left[\ln\left(\frac{\Phi}{b}\right)\right]^2
\left(-4^{\frac{1}{1-k}}k^2\Phi^{\frac{2k}{k-1}}+\Phi^2k(k-1)\right)\\
&&-4k(k-1)(d-1)\ln\left(\frac{\Phi}{b}\right)\left(4^{\frac{1}{1-k}}\Phi^{\frac{2k}{k-1}}-
\Phi^2\right)+(d-2)(d-1)(k-1)\left(-4^{\frac{1}{1-k}}(k-1)\Phi^{\frac{2k}{k-1}}+\Phi^2 k\right)\Big\}.\nonumber
\end{eqnarray}
and the equations of motion (\ref{eqsmotionk}) admit a solution where the scalar field and the metric are given as follows
\begin{eqnarray}
\Phi(r)=b \,e^{ar}, \quad F(r)=r^2\left(1-\left[\frac{1}{4}\, b^2\, e^{2ar} \right]^{\frac{1}{k-1}}\right) \, .
\nonumber
\end{eqnarray}

\section{Thermodynamics}
The partition function for a thermodynamical ensemble is identified
with the Euclidean path integral in the saddle point approximation
around the Euclidean continuation of the classical solution
\cite{Gibbons:1976ue}. The Euclidean and Lorentzian action are
related by $I_{E}=-iI$ where the periodic Euclidean time is  $\tau
=it$. The Euclidean continuation of the class of metrics considered here  is given by
$$
ds^2=N(r)^2F(r)d\tau^2+\frac{dr^2}{F(r)}+r^2\left(dx_1^2+dx_2^2+\cdots+dx_{d-2}^2\right).
$$
In order to avoid conical singularity at the horizon in the
Euclidean metric, the Euclidean time is made periodic with period
$\beta$ and the Hawking temperature $T$ is given by $T=\beta^{-1}$.
Here we are interested only in the static solution with a radial scalar field, and hence it is enough to consider a
\textit{reduced} action principle. This latter is given by
\begin{eqnarray}
\label{redaction}
I_E=&&-\beta\Sigma_{d-2}\int_{r_h}^{\infty}\frac{N}{2k}(d-2)\frac{d}{dr}\left[r^{d-1}\left(1-\frac{F}{r^2}\right)^k\right]\nonumber\\&&+
\beta\Sigma_{d-2}\int_{r_h}^{\infty} N r^{d-2}\Big\{\left(\frac{1-4\xi}{2}\right)F(\Phi^{\prime})^2-
\xi\Phi\Phi^{\prime}\left(F^{\prime}+\frac{2(d-2)}{r}F\right)-2\xi\Phi\Phi^{\prime\prime}F\nonumber\\&&-\frac{(d-2)\xi}{2r}F^{\prime}\Phi^2
+\Phi^2\left(-\frac{\xi}{2r^2}(d-2)(d-3)F\right)+U(\Phi)\Big\}+B
\end{eqnarray}
where the radial coordinate $r$ belongs to the range $[r_h,\infty[$ where $r_h$ is the location of the horizon. Here,
$\beta$ stands for the inverse of the temperature and $\Sigma_{d-2}$ corresponds
to the compactified volume of the planar $(d-2)-$dimensional base manifold.
In the reduced action (\ref{redaction}), $B$ is a boundary term that is fixed by requiring that the Euclidean action has an extremum, that is $\delta I_{E}=0$
which in turn implies that
\begin{eqnarray}
{\delta B}=\beta\Sigma_{d-2}\left({\delta B}_{\mbox{\tiny{gravity}}}+{\delta B}_{\mbox{\tiny{matter}}}\right)
\end{eqnarray}
where the first contribution ${\delta B}_{\mbox{\tiny{gravity}}}$ arises from the variation of the gravity part while the second one
${\delta B}_{\mbox{\tiny{matter}}}$ comes from the matter source, and are given by
\begin{eqnarray}
{\delta B}_{\mbox{\tiny{gravity}}}=&&\int_{r_h}^{\infty}\frac{N}{2k}(d-2)\frac{d}{dr}\left[r^{d-1}\delta\left(1-\frac{F}{r^2}\right)^k\right],\nonumber\\
\\
{\delta B}_{\mbox{\tiny{matter}}}=&&\int_{r_h}^{\infty}2\xi\Phi Fr^{d-2}\delta\Phi^{\prime}+\int_{r_h}^{\infty}\delta F\left[\frac{\xi(d-2)}{2}r^{d-3}\Phi^2+\xi\Phi\Phi^{\prime}r^{d-2}\right]\nonumber\\
&&+\int_{r_h}^{\infty}\Big\{\delta\Phi\left[F\Phi^{\prime}r^{d-2}(2\xi-1)-\xi F^{\prime}\Phi r^{d-2}\right]\Big\}.\nonumber
\end{eqnarray}
In the grand canonical ensemble, the Euclidean action is related with the mass ${\cal M}$ and the entropy ${\cal S}$ by
\begin{eqnarray}
I_E=\beta {\cal M}-{\cal S}.
\label{gce}
\end{eqnarray}
For the solution (\ref{solk}), we have the following asymptotic variation behaviors
\begin{eqnarray*}
&&\delta F|_{\infty}=-\alpha \xi^{\frac{1}{k-1}}a^{\alpha-1}(\delta a)\left[r^{\alpha+2}+\frac{b(\alpha-1)}{a}r^{\alpha+1}
+\frac{b^2(\alpha-1)(\alpha-2)}{a^2}r^{\alpha}+\cdots\right],\nonumber\\
&&\delta\Phi|_{\infty}=a^{\frac{(k-1)\alpha}{2}}(\delta a)\left[r^{\frac{(k-1)\alpha}{2}}+\frac{(k-1)\alpha b}{2a}r^{\frac{(k-1)\alpha-2}{2}}+\frac{(k-1)\alpha\left((k-1)\alpha-2\right)b^2}{4a^2}r^{\frac{(k-1)\alpha-4}{2}}
+\cdots\right],\\
&&\delta\Phi^{\prime}|_{\infty}=\frac{d}{dr}\left(\delta\Phi|_{\infty}\right),\nonumber
\label{inf}
\end{eqnarray*}
where for simplicity we have defined $\alpha=\frac{4\xi}{(k-1)(4\xi-1)}$. At the horizon $r_h$ the variations read
\begin{eqnarray*}
\delta F|_{r_h}=-F^{\prime}|_{r_h}\delta r_h,\qquad
\delta\Phi|_{r_h}=\delta\Phi(r_h)-\Phi^{\prime}|_{r_h}\delta r_h
\label{hor}
\end{eqnarray*}
It is intriguing that for the solution (\ref{solk}), the variation of the gravity part at the infinity  exactly cancels at each order the
variation of the matter at infinity, that is
$$
{\delta B}_{\mbox{\tiny{gravity}}}(\infty)=-{\delta B}_{\mbox{\tiny{matter}}}(\infty),
$$
and hence we have $B(\infty)=0$. At the horizon, a simple computation yields
$$
\delta B(r_h)=2\pi\Sigma_{d-2}\delta\Big[\left(1-\xi\,\Phi^2(r_h)\right)r_h^{d-2}\Big]
$$
but as stressed before (\ref{truc}) the quantity between brackets vanishes. Hence,
the boundary term vanishes identically, $B=0$, and  the
identification of the mass and the entropy through (\ref{gce}) yields
$$
{\cal M}=0,\qquad {\cal S}=0.
$$
It is then clear that the solutions obtained here have the particularity that their entropy is proportional to the lapse function
evaluated at the horizon (\ref{truc}), and this quantity, by definition of the horizon, precisely vanishes. Consequently, the integration constant $a$ appearing in the solution can be naively interpreted as a sort of hair since it has not conserved charged associated to it. The temperature of the solution is given by
\begin{eqnarray}
T=-\frac{\xi^{\frac{8\xi-1}{4\xi}}}{\pi a (k-1)(4\xi-1)}\left(\xi^{\frac{1-4\xi}{4\xi}}-b\right)^2,
\end{eqnarray}
and turns out to be positive provided that the nonminimal coupling parameter $\xi<\frac{1}{4}$ which is precisely the range where the solutions behaves asymptotically AdS. Interestingly enough, the temperature being inversely proportional to the hair $a$, this implies that a high temperature the hair will disappear as it is excepted in order to reproduce the phase diagram of the unconventional superconductors.

\section{Conclusions}
Here, we have considered some particular Lovelock gravity theories indexed by an integer $k$ whose coefficients are fixed by requiring the existence
of a unique AdS vacuum with a matter source given by a self-interacting nonminimally coupled scalar field. We have shown that for each inequivalent Lovelock gravity theories there exists an appropriate choice for the self-interacting potential that permits to obtain black hole solutions for any arbitrary values of the nonminimal coupling parameter. The form of the potential involves six different terms where the first three terms exactly
correspond to the potential that usually arises in different situations involving nonminimally coupled scalar fields. It will be interesting to explore in which context the $k-$depending part of the potential $U_k$ (which concerns the last three terms in (\ref{potLov})) may emerge. The thermodynamics analysis of the solutions shows that the mass and the entropy of the solutions both vanish, and hence the integration constant $a$ of the solutions can be viewed as a sort of hair.  Note that in the context of pure Lovelock gravity in even dimension, black hole solutions with zero mass and zero entropy have been found in \cite{Anabalon:2011bw}. In our case, the vanishing of the thermodynamical quantities can be viewed as a consequence of the fact that the entropy of the solutions are proportional to the lapse function evaluated at the horizon (\ref{truc}). We have shown that the temperature $T$ goes like $T\propto a^{-1}$ and hence at high temperature the hair should disappear as it is excepted in  order to reproduce the phase diagram of the unconventional superconductors. It will be more than interesting to explore more intensively the possible applications of the solutions derived here in the context of holographic superconductors. We end with the fact that for a coupling constant $b=0$ and $\xi=\hat{\xi}_{k,d+1}$ as defined in (\ref{xi}), the potential reduces to a mass term and the solution becomes a stealth configuration \cite{Gaete:2013ixa,Gaete:2013oda}, that is a particular solution of the field equations (\ref{eqsmotionk}) where both sides of the equations  (gravity and matter)  vanish identically
$$
{\cal G}^{(k)}=0=T_{\mu\nu}.
$$
In this particular case, the solution involves an additional integration constant which is due to the fact that the equation $T_{\mu\nu}=0$ becomes invariant under the rescaling of the scalar field $\Phi\to A\Phi$, where $A$ is a constant. In order to be self-contained, we present the stealth solution as found in \cite{Gaete:2013ixa,Gaete:2013oda} for $b=0$ and $\xi=\hat{\xi}_{k,d+1}$
\begin{eqnarray}
\label{ss}
&&ds^{2}=-\left(r^2-(ar)^{\frac{1-d}{k}}\right)+\frac{dr^2}{\left(r^2-(ar)^{\frac{1-d}{k}}\right)}+{r^{2}}d\vec{x}_{d-2}^2,\nonumber\\
&&\Phi=Ar^{\frac{(k-1)(1-d)}{2k}}.\nonumber
\end{eqnarray}
Because of this additional constant $A$, the quantity $(1-\xi\Phi^2)$ evaluated at the horizon may not necessarily be zero, and consequently,
the entropy may not be zero. We would like to further explore this issue concerning the thermodynamics properties of the stealth solutions arising in the context of nonminimally coupled scalar field.

\begin{acknowledgments}
We thank Moises Bravo, Julio Oliva, Alfredo Perez and Ricardo Troncoso for useful discussions. FC is partially supported by the Fondecyt grant 11121651 and by the Conicyt grant  79112034. MH is partially
supported by grant 1130423 from FONDECYT, by grant ACT 56 from
CONICYT and from CONICYT, Departamento de Relaciones Internacionales
``Programa Regional MATHAMSUD 13 MATH-05''. The Centro de Estudios Cient\'{\i}ficos (CECs) is funded by the Chilean Government
through the Centers of Excellence Base Financing Program of Conicyt.
\end{acknowledgments}



\begin{thebibliography}{99}



\bibitem{LAN} C. Lanczos, ``A remarkable property of the Riemann.Christoffel tensor in four dimensions,'' Ann. Math. {\bf 39}, 842 (1938).

\bibitem{LOV} D. Lovelock, ``The Einstein tensor and its generalizations,'' J. Math. Phys {\bf 12}, 498 (1971).



\bibitem{Zanelli:2005sa}
  J.~Zanelli,
  ``Lecture notes on Chern-Simons (super-)gravities. Second edition (February 2008),''
  hep-th/0502193.



\bibitem{Banados:1993ur}
  M.~Banados, C.~Teitelboim and J.~Zanelli,
  ``Dimensionally continued black holes,''
  Phys.\ Rev.\ D {\bf 49}, 975 (1994)
  [gr-qc/9307033].

\bibitem{Cai:1998vy}
  R.~-G.~Cai and K.~-S.~Soh,
  ``Topological black holes in the dimensionally continued gravity,''
  Phys.\ Rev.\ D {\bf 59}, 044013 (1999)
  [gr-qc/9808067].

\bibitem{Crisostomo:2000bb}
  J.~Crisostomo, R.~Troncoso and J.~Zanelli,
  ``Black hole scan,''
  Phys.\ Rev.\ D {\bf 62}, 084013 (2000)
  [hep-th/0003271].

\bibitem{Aros:2000ij}
  R.~Aros, R.~Troncoso and J.~Zanelli,
  ``Black holes with topologically nontrivial AdS asymptotics,''
  Phys.\ Rev.\ D {\bf 63}, 084015 (2001)
  [hep-th/0011097].


\bibitem{Boulware:1985wk}
  D.~G.~Boulware and S.~Deser,
  ``String Generated Gravity Models,''
  Phys.\ Rev.\ Lett.\  {\bf 55}, 2656 (1985).


\bibitem{Cai:2001dz}
  R.~-G.~Cai,
"Gauss-Bonnet black holes in AdS spaces,''
  Phys.\ Rev.\ D {\bf 65}, 084014 (2002)
  [hep-th/0109133].

















\bibitem{Bekenstein:1974sf}
  J.~D.~Bekenstein,
  ``Exact Solutions Of Einstein Conformal Scalar Equations,''
  Annals Phys.\  {\bf 82}, 535 (1974).
\bibitem{Bocharova:1970}
N.~M.~Bocharova, K.~A.~Bronnikov and V.~N.~Melnikov, ``An exact
solution of the system of Einstein equations and mass-free scalar
field,'' Vestn. Mosk. Univ. Fiz. Astron. {\bf 6}, 706 (1970)  [Moscow
Univ. Phys. Bull. {\bf 25}, 80 (1970)].






\bibitem{Martinez:2002ru}
  C.~Martinez, R.~Troncoso and J.~Zanelli,
  ``De Sitter black hole with a conformally coupled scalar field in  four
  dimensions,''
  Phys.\ Rev.\  D {\bf 67}, 024008 (2003)
  [arXiv:hep-th/0205319].
\bibitem{Martinez:2005di}
  C.~Martinez, J.~P.~Staforelli and R.~Troncoso,
  ``Charged topological black hole with a conformally coupled scalar field,''
  Phys.\ Rev.\  D {\bf 74}, 044028 (2006)
  [arXiv:hep-th/0512022].




\bibitem{Horowitz:2010gk}
  G.~T.~Horowitz,
  ``Introduction to Holographic Superconductors,''
  arXiv:1002.1722 [hep-th].

\bibitem{Hartnoll:2008kx}
  S.~A.~Hartnoll, C.~P.~Herzog and G.~T.~Horowitz,
  ``Holographic Superconductors,''
  JHEP {\bf 0812}, 015 (2008)

\bibitem{Herzog:2009xv}
  C.~P.~Herzog,
  ``Lectures on Holographic Superfluidity and Superconductivity,''
  J.\ Phys.\ A {\bf 42}, 343001 (2009) 
  [arXiv:0904.1975 [hep-th]].



\bibitem{Gaete:2013ixa}
  M. Bravo-Gaete and M.~Hassaine,
  ``Topological black holes for Einstein-Gauss-Bonnet gravity with a nonminimal scalar field,''
  Phys.\ Rev.\ D {\bf 88}, 104011 (2013)
  [arXiv:1308.3076 [hep-th]].


\bibitem{Gaete:2013oda}
  M. Bravo-Gaete  and M.~Hassaine,
  ``Planar AdS black holes in Lovelock gravity with a nonminimal scalar field,''
  JHEP {\bf 1311}, 177 (2013)
  [arXiv:1309.3338 [hep-th]].







\bibitem{AyonBeato:2004ig}
  E.~Ayon-Beato, C.~Martinez and J.~Zanelli,
  ``Stealth scalar field overflying a (2+1) black hole,''
  Gen.\ Rel.\ Grav.\  {\bf 38}, 145 (2006)
  [hep-th/0403228].


\bibitem{Banados:1992wn}
  M.~Banados, C.~Teitelboim and J.~Zanelli,
  ``The Black hole in three-dimensional space-time,''
  Phys.\ Rev.\ Lett.\  {\bf 69}, 1849 (1992)
  [hep-th/9204099].


\bibitem{AyonBeato:2005qq}
  E.~Ayon-Beato and M.~Hassaine,
  ``Exploring AdS waves via nonminimal coupling,''
  Phys.\ Rev.\ D {\bf 73}, 104001 (2006)
  [hep-th/0512074].




\bibitem{Caldarelli:2013gqa}
  M.~M.~Caldarelli, C.~Charmousis and M.~Hassaine,
  ``AdS black holes with arbitrary scalar coupling,''
  JHEP {\bf 1310}, 015 (2013)
  [arXiv:1307.5063 [hep-th]].


\bibitem{Gibbons:1976ue}
  G.~W.~Gibbons and S.~W.~Hawking,
  ``Action Integrals and Partition Functions in Quantum Gravity,''
  Phys.\ Rev.\ D {\bf 15}, 2752 (1977).








































\bibitem{Anabalon:2011bw}
  A.~Anabalon, F.~Canfora, A.~Giacomini and J.~Oliva,
  ``Black holes with gravitational hair in higher dimensions,''
  Phys.\ Rev.\ D {\bf 84}, 084015 (2011)
  [arXiv:1108.1476 [hep-th]].

\end{thebibliography}
\end{document}